\documentclass[aps,prb,twocolumn,showpacs]{revtex4}
\usepackage{mathrsfs}
\usepackage{amssymb}
\usepackage{amsmath}
\usepackage{graphicx}
\usepackage{dcolumn}
\usepackage{bm}
\begin{document}

\title{Abelian and Non-Abelian Quantum Geometric Tensor}

\author{Yu-Quan Ma}
\email{mayuquan@iphy.ac.cn} \affiliation{Beijing National Laboratory
for Condensed Matter Physics, Institute of Physics, Chinese Academy
of Sciences, Beijing 100190, China}

\author{Shu Chen}
\affiliation{Beijing National Laboratory for Condensed Matter
Physics, Institute of Physics, Chinese Academy of Sciences, Beijing
100190, China}

\author{Heng Fan}
\affiliation{Beijing National Laboratory for Condensed Matter
Physics, Institute of Physics, Chinese Academy of Sciences, Beijing
100190, China}

\author{Wu-Ming Liu}
\affiliation{Beijing National Laboratory for Condensed Matter
Physics, Institute of Physics, Chinese Academy of Sciences, Beijing
100190, China}

\begin{abstract}
We propose a generalized quantum geometric tenor to understand
topological quantum phase transitions, which can be defined on the
parameter space with the adiabatic evolution of a quantum many-body
system. The generalized quantum geometric tenor contains two
different local measurements, the non-Abelian Riemannian metric and
the non-Abelian Berry curvature, which are recognized as two natural
geometric characterizations for the change in the ground-state
properties when the parameter of the Hamiltonian varies. Our results
show the symmetry-breaking and topological quantum phase transitions
can be understood as the singular behavior of the local and
topological properties of the quantum geometric tenor in the
thermodynamic limit.
\end{abstract}

\maketitle

\section{Introduction.}

Quantum phase transitions (QPTs) are characterized by the
\emph{qualitative} changes in the ground-state properties when the
parameter of a quantum many-body system varies. Different from
classical phase transitions driven by the thermal fluctuations, QPTs
occur at zero temperature and thus are driven entirely by the
quantum fluctuations. \cite{Sachdev} Traditionally, phase
transitions are understood in the Landau-Ginzburg-Wilson paradigm
with symmetry breaking and local order parameter. However, it has
been revealed that this paradigm may not describe all possible
orders and a new one \emph{topological order} \cite{Wen} needs to
come into physics since the study of the integer and fractional
quantum Hall states. Whereas the preexistent symmetry in the
Hamiltonian is absent, the phases can only be distinguished by
certain topological invariants associated with the robust
degeneracies of the ground-state instead of any local order
parameter. Historically, this was first pointed out in the integer
quantum Hall state and the corresponding topological invariant is
recognized as the first Chern number. \cite{TKNN,Simon,Niu} Up to
now, topological orders can be partially described by a new set of
quantum numbers, such as ground-state degeneracy, quasi-particle
fractional statistics, edge states and topological entropy.
\cite{Kitaev-Levin} However, the classification of topological
orders is still an open question.

More recently, the concept of quantum geometric tensor (QGT) (Ref.
7) based on differential-geometry has been introduced to analyze
continuous QPTs. \cite{QGT-Zanardi,Venuti} In this framework, the
two approaches of the ground-state Berry phase and fidelity to
witness QPTs are unified. It is shown that the underlying mechanism
is the singular behavior of the QGT in the vicinity of the critical
points. Specially, the real part of the QGT is a Riemannian metric
defined over the parameter manifold, while the imaginary part is the
Berry curvature which flux give rise to the Berry phase. The
Riemannian metric is recognized as the leading term of the fidelity
\cite{Zanardi} which is the overlap of two ground states associated
to infinitesimally close parameters. Generally, the Riemannian
metric will exhibit the divergent behavior in the quantum critical
region in the thermodynamic limit. In the approach of Berry phase,
it was argued that a non-contractible ground-state Berry phase in
the loop over the parameter space is associated to QPTs.
\cite{Hamma} This fact indicates that the critical points associated
to the divergence of the Berry curvature in the thermodynamic limit.
\cite{Mayuquan} Particularly, a scaling analysis of this QGT in the
vicinity of the critical points has been performed. \cite{Venuti} So
far, the approach of QGT has been applied to detect the phases
boundaries in various systems. \cite{FS}

What is underlying physical mechanism which endow the QGT with the
Berry curvature and can it be extended to the non-Abelian case?
Whether the QGT could be extended to a topological invariant and to
distinguish the different topological phases not only the phases
boundaries? In this paper, we present an adiabatic origin for the
QGT and show a generalized non-Abelian QGT will naturally emerge
from characterizing the distance between two neighbor states in the
$U(n)$ vector bundle induced by the quantum adiabatic evolution of
the degenerate system. In addition to a non-Abelian Riemannian
metric as its symmetric part, the generalized QGT has also been
equipped with a non-Abelian Berry curvature as its anti-symmetric
part. We demonstrate that this is due to the reduction in the
parallel transport of states with the adiabatic evolution.
Furthermore, we find its symmetric and anti-symmetric parts
correspond to two different local geometric measurements, which
provide intrinsic characterizations for the change degree of the
ground-state properties when the parameter of the Hamiltonian
varies. As a general gauge covariant, the generalized QGT can be
extended to the topological invariants Chern numbers because its
anti-symmetric part is just the instinct curvature tensor of the
induced fiber bundles. This approach may provide a single framework
to analyze a class of QPTs with topological orders.

\section{Adiabatic Origin of the Abelian Quantum Geometric
Tensor}

Let us consider a family of parameter-dependent Hamiltonian
$\{H(\lambda )\}$ for a quantum many-body system. We require
$H(\lambda )$ to depend smoothly on a set of parameters $\lambda
=\left( \lambda ^{1},\lambda ^{2},\cdots \right) \in \mathcal{M}$
($\mathcal{M}$ denotes the Hamiltonian parameters base manifold) and
act over the Hilbert-space $\mathcal{H}(\lambda )$.
Especially, we are interested in the properties of the ground-state $%
\left\vert {\Psi _{0}\left( \lambda \right) }\right\rangle $ of the
eigenvalue $E_{0}(\lambda )$. To begin with, we restrict our discussion on
the $U\left( 1\right) $ case for convenience. Therefore the eigenvalue $%
E_{0}(\lambda )$ is required non-degenerate about all the parameters over $%
\mathcal{M}$, that is, the eigenspace $\mathcal{H}_{E_{0}}(\lambda )$ of $%
E_{0}(\lambda )$ is required to be one-dimensional. Now we associate
each $\lambda $ with a complex vector space
$\mathcal{H}_{E_{0}}(\lambda )$
and thus forms a $U{(1)}$ line bundle. For the ground-state energy $%
E_{0}(\lambda )$, the subspace $\mathcal{H}_{E_{0}}(\lambda ):=$Span$%
\{\left\vert {\psi }_{0}{\left( \lambda \right) }\right\rangle \}$, where $%
\left\vert {\psi }_{0}{\left( \lambda \right) }\right\rangle $ is a complete
orthonormal basis. In this basis, the ground-state can be written as
\begin{equation}
\left\vert {\Psi _{0}\left( \lambda \right) }\right\rangle ={C\left( \lambda
\right) }\left\vert {\psi }_{0}{\left( \lambda \right) }\right\rangle .
\label{wavefunc}
\end{equation}%
Obviously, the choice of basis is quite arbitrary and the component ${C}%
\left( \lambda \right) $ ($|{C}\left( \lambda \right) |^{2}=1$) of the
ground-state $\left\vert {\Psi _{0}\left( \lambda \right) }\right\rangle $
in different basis is related by a $U\left( \mathrm{1}\right) $ local gauge
transformation. 
Consider the distance between two neighbor states over the parameters base
manifold $\mathcal{M}$, we have $dS:=\left\Vert {\Psi }_{0}{\left( {\lambda
+\delta \lambda }\right) -\Psi }_{0}\left( {\lambda }\right) \right\Vert $,
which can be expanded to
\begin{equation}
dS^{2}=\sum_{\mu ,\upsilon }{\left\langle {{\partial _{\mu }\Psi }}_{0}{%
\left( \lambda \right) {d\lambda ^{\mu }}}\mathrel{\left | {\vphantom
{{\partial _\mu \Psi_{0} d\lambda ^\mu } {\partial _\upsilon \Psi_{0}
d\lambda ^\upsilon }}} \right. \kern-\nulldelimiterspace}{{\partial
_{\upsilon }\Psi _{0}d\lambda ^{\upsilon }}}{\left( \lambda \right) }%
\right\rangle },  \label{1ds2}
\end{equation}%
where $\partial _{\mu }\Psi _{0}{\left( \lambda \right) }:={\partial \Psi
_{0}\left( \lambda \right) }/{\partial \lambda ^{\mu }}$. By using Eq. (\ref%
{wavefunc}), the term $\left\vert {\partial _{\mu }\Psi _{0}\left( \lambda
\right) }\right\rangle $ can be expanded as $\left\vert {\partial _{\mu
}\Psi _{0}\left( \lambda \right) }\right\rangle ={{\partial _{\mu }C\left(
\lambda \right) \cdot \left\vert {\psi }_{0}{\left( \lambda \right) }%
\right\rangle +C{\left( \lambda \right) }\left\vert {\partial _{\mu }\psi }%
_{0}{\left( \lambda \right) }\right\rangle .}}$ Note that the term ${{C{%
\left( \lambda \right) }\left\vert {\partial _{\mu }\psi }_{0}{\left(
\lambda \right) }\right\rangle }}$ will generally be mapped out of the
sub-space $\mathcal{H}_{E_{0}}(\lambda )$. Therefore we introduce the
sub-space $\mathcal{H}_{E_{0}}(\lambda )$ projection operator $\mathcal{P}%
(\lambda ):=\left\vert {\psi _{0}\left( \lambda \right) }\right\rangle
\left\langle {\psi _{0}\left( \lambda \right) }\right\vert $ and the
complete projection operator $\boldsymbol{1}$. Then we have
\begin{eqnarray}
\left\vert {\partial _{\mu }\Psi _{0}}(\lambda )\right\rangle &=&\left[
\partial _{\mu }C{\left( \lambda \right) }\left\vert {\psi _{0}\left(
\lambda \right) }\right\rangle +C{\left( \lambda \right) }\mathcal{P}%
(\lambda )\left\vert {\partial _{\mu }\psi _{0}\left( \lambda \right) }%
\right\rangle \right]  \notag \\
&&+C{\left( \lambda \right) }\left[ \boldsymbol{1-}\mathcal{P}(\lambda )%
\right] \left\vert {\partial _{\mu }\psi _{0}\left( \lambda \right) }%
\right\rangle .  \label{pianwave}
\end{eqnarray}

Now we introduce the quantum adiabatic evolution to the system and suppose
that the $\mathrm{dim}\mathcal{H}_{E_{0}}(\lambda )$ does not depend on the
parameter $\lambda $, and particularly, demand that during the evolution of
the system there are no level crossings. Under the quantum adiabatic limit,
the evolution of the ground-state $\left\vert {\Psi _{0}\left( \lambda
\right) }\right\rangle \in \mathcal{H}_{E_{0}}(\lambda )$ to $\left\vert {%
\Psi _{0}\left( \lambda +\delta \lambda \right) }\right\rangle \in \mathcal{H%
}_{E_{0}}(\lambda +\delta \lambda )$ will undergo a parallel transport in
the sense of Levi-Civit\`{a} from $\lambda $ to $\lambda +\delta \lambda $
on the base manifold $\mathcal{M}$. We have
\begin{equation}
\left\vert {D_{\mu }\Psi _{0}}(\lambda )\right\rangle =0,  \label{covariant}
\end{equation}%
where $\left\vert {D_{\mu }\Psi _{0}}(\lambda )\right\rangle $ denote the
covariant derivative of $\left\vert {\Psi _{0}}(\lambda )\right\rangle $ in
the ${U}\left( 1\right) $ line bundle, which can be written in the basis of $%
\{\left\vert {\psi }_{0}{\left( \lambda \right) }\right\rangle \}$ as $%
\left\vert {D_{\mu }\Psi _{0}}(\lambda )\right\rangle =\partial _{\mu }C{%
\left( \lambda \right) }\cdot \left\vert {\psi _{0}\left( \lambda \right) }%
\right\rangle +C{\left( \lambda \right) }\cdot \mathcal{P}(\lambda )\cdot
\left\vert {\partial _{\mu }\psi _{0}\left( \lambda \right) }\right\rangle $%
. It is clear that the term of $\mathcal{P}(\lambda )\cdot \left\vert {%
\partial _{\mu }\psi _{0}\left( \lambda \right) }\right\rangle $ play the
role of connection. We have $\mathcal{P}(\lambda )\cdot \left\vert {\partial
_{\mu }\psi _{0}\left( \lambda \right) }\right\rangle =-i\mathcal{A}_{\mu
}\left\vert {\psi _{0}\left( \lambda \right) }\right\rangle $, where the
connection coefficients $\mathcal{A}_{\mu }=:i\langle \psi _{0}|\partial
_{\mu }\psi _{0}\rangle $ is just the Berry connection. Substituting Eqs. (%
\ref{pianwave}) and (\ref{covariant}) to Eq. (\ref{1ds2}), we have $%
\langle {\partial _{\mu }\Psi _{0}}(\lambda )|{\partial _{{\nu }}\Psi _{0}}%
(\lambda )\rangle =\langle {\partial _{\mu }\psi _{0}\left( \lambda \right) }%
|\left[ \boldsymbol{1-}\mathcal{P}(\lambda )\right] \left\vert {\partial _{{%
\nu }}\psi _{0}\left( \lambda \right) }\right\rangle $. Thus we get a
Hermitian metric tensor
\begin{equation}
Q_{\mu {\nu }}:=\langle {\partial _{\mu }\psi _{0}\left( \lambda \right) }|%
\left[ \boldsymbol{1-}\mathcal{P}(\lambda )\right] \left\vert {\partial _{{%
\nu }}\psi _{0}\left( \lambda \right) }\right\rangle .  \label{QGT1}
\end{equation}%
This is the adiabatic origin of the Abelian quantum geometric tensor. The
Hermitian metric $Q_{\mu {\nu }}$ defines a positive definite symmetric
Riemannian metric $g_{\mu {\nu }}:=\left( Q_{\mu {\nu }}+Q_{\mu {\nu }%
}^{\ast }\right) /2$ $\left[\text{in the }U(1)\text{ case, }g_{\mu {\nu }%
}=\Re Q_{\mu {\nu }}\right] $, and then the distance can be written
as
\begin{equation}
dS^{2}=\sum_{\mu ,\upsilon }g_{\mu {\nu }}{d\lambda ^{\mu }d\lambda ^{\nu }.}
\label{ds2}
\end{equation}%
We can also associate to $Q_{\mu {\nu }}$ a $2$-form $F=\sum_{\mu ,\upsilon
}F_{\mu {\nu }}{d\lambda ^{\mu }\wedge d\lambda ^{\nu }}$, where $F_{\mu {%
\nu }}:=i\left( Q_{\mu {\nu }}-Q_{\mu {\nu }}^{\ast }\right) $ $\left[ \text{%
in the }U(1)\text{ case, }F_{\mu {\nu }}=-2\Im Q_{\mu {\nu
}}\right]$, which is nothing but the Berry curvature. Furthermore we
can get the relation
\begin{equation}
Q_{\mu {\nu }}=g_{\mu {\nu }}-\frac{i}{2}F_{\mu {\nu }}.  \label{QGT2}
\end{equation}

\section{Non-Abelian Quantum Geometric Tensor}

Here we point out that the QGT can be extended to a more general
case in which a non-Abelian QGT can be naturally defined on the
$U(N)$ vector bundle induced by the adiabatic development of the
quantum degenerate system. This approach is
very similar to the $U(1)$ case but now the sub-space $\mathcal{H}%
_{E_{0}}(\lambda )$ is a $ND$ Hilbert-space. For the ground energy $%
E_{0}(\lambda )$, the eigenspace is $\mathcal{H}_{E_{0}}(\lambda ):=$Span$%
\{\left\vert {\psi }_{i}{\left( \lambda \right) }\right\rangle \}_{i=1}^{N}$%
. A ground-state $\left\vert {\Psi _{0}\left( \lambda \right) }\right\rangle
\in \mathcal{H}_{E_{0}}(\lambda )$ can be expanded as $\left\vert {\Psi
_{0}\left( \lambda \right) }\right\rangle =\sum_{i=1}^{N}{C_{i}}\left(
\lambda \right) \left\vert {\psi }_{i}{\left( \lambda \right) }\right\rangle
$ and the wave function $\left[ {C_{1}}\left( \lambda \right) ,{C_{2}}\left(
\lambda \right) ,\cdots ,C_{N}\left( \lambda \right) \right] ^{T}$
represented in different basis is related by a $U\left( N\right) $ local
gauge transformation. The distance between two neighbor states over $%
\mathcal{M}$ is $dS^{2}=\sum_{\mu ,\upsilon }{\left\langle {{\partial _{\mu
}\Psi }}_{0}{\left( \lambda \right) {d\lambda ^{\mu }}}%
\mathrel{\left | {\vphantom
{{\partial _\mu \Psi_{0} d\lambda ^\mu } {\partial _\upsilon
\Psi_{0} d\lambda ^\upsilon }}} \right. \kern-\nulldelimiterspace}{{\partial
_{\upsilon }\Psi _{0}d\lambda ^{\upsilon }}}{\left( \lambda \right) }%
\right\rangle }$, in which the term $\left\vert {\partial _{\mu }\Psi
_{0}\left( \lambda \right) }\right\rangle $ can be decomposed as $\left\vert
{\partial _{\mu }\Psi _{0}\left( \lambda \right) }\right\rangle =\left\vert {%
D_{\mu }\Psi _{0}}(\lambda )\right\rangle {{+\sum_{i=1}^{N}{{C_{i}}\left(
\lambda \right) }}}\left[ \boldsymbol{1}-\mathcal{P}(\lambda )\right] {{%
\left\vert {\partial _{\mu }\psi }_{i}{\left( \lambda \right) }\right\rangle
}}$, where $\mathcal{P}(\lambda ):=\oplus _{j=1}^{N}\left\vert {\psi
_{j}\left( \lambda \right) }\right\rangle \left\langle {\psi _{j}\left(
\lambda \right) }\right\vert $ is the projection operator and $\left\vert {%
D_{\mu }\Psi _{0}}(\lambda )\right\rangle =\sum_{i=1}^{N}\partial _{\mu
}C_{i}{\left( \lambda \right) }\cdot \left\vert {\psi _{i}\left( \lambda
\right) }\right\rangle +C_{i}{\left( \lambda \right) }\cdot \mathcal{P}%
(\lambda )\cdot \left\vert {\partial _{\mu }\psi _{i}\left( \lambda \right) }%
\right\rangle $ denote the covariant derivative of $\left\vert {\Psi _{0}}%
(\lambda )\right\rangle $ in the ${U}\left( n\right) $ vector bundle.

Now let us introduce the quantum adiabatic evolution to the degenerate
system, therefore $\left\vert {D_{\mu }\Psi _{0}}(\lambda )\right\rangle =0$%
, which means the ground state $\left\vert {\Psi _{0}\left( \lambda \right) }%
\right\rangle $ changing to $\left\vert {\Psi _{0}\left( \lambda
+\Delta \lambda \right) }\right\rangle $ along an arbitrary path
$\Gamma \left( \lambda \right) $ on the parameter base manifold
$\mathcal{M}$ will undergo a parallel transport. Note that the term
$\mathcal{P}(\lambda )\cdot \left\vert {\partial _{\mu }\psi
_{m}\left( \lambda \right) }\right\rangle $
$=-i\sum_{l=1}^{N}\mathcal{A}_{\mu }^{lm}\left\vert {\psi _{l}\left(
\lambda \right) }\right\rangle $, here $\mathcal{A}_{\mu
}^{lm}=i\langle \psi _{l}|\partial _{\mu }\psi _{m}\rangle $ is
known as the Wilczek-Zee connection, \cite{Wil-Zee} which is a
matrix element of the non-Abelian
gauge potential. 
By using the above relations, we can derive the distance between two
neighbor states over $\mathcal{M}$ as
\begin{eqnarray}
dS^{2} &=&\sum_{\mu ,\upsilon }{\left\langle {{\partial _{\mu }\Psi }}_{0}{%
\left( \lambda \right) {d\lambda ^{\mu }}}\mathrel{\left | {\vphantom
{{\partial _\mu \Psi_{0} d\lambda ^\mu } {\partial _\upsilon \Psi_{0}
d\lambda ^\upsilon }}} \right. \kern-\nulldelimiterspace}{{\partial
_{\upsilon }\Psi _{0}d\lambda ^{\upsilon }}}{\left( \lambda \right) }%
\right\rangle }  \notag \\
=\!\sum\limits_{\mu \nu }\!\!\!\!\!\!\!\!\!\!\!\! &&{\left[ {\left( {%
\begin{array}{*{20}c} {C^* _1 } & \cdots & {C^* _n } \\ \end{array}}\right) {%
Q_{\mu \upsilon }}\left( {\begin{array}{*{20}c} {C_1 } \\ \vdots \\ {C_n }
\\ \end{array}}\right) }\right] }{d\lambda ^{\mu }d\lambda ^{\upsilon },}
\label{nads2}
\end{eqnarray}%
where ${Q_{\mu \upsilon }}$ is a $N\times N$ Hermitian matrix. We define
this quantity as the non-Abelian QGT. Its matrix element is given by
\begin{equation}
{{Q_{\mu \upsilon }^{ij}}}:=\langle {\partial _{\mu }\psi _{i}\left( \lambda
\right) }|\left[ \boldsymbol{1-}\mathcal{P}(\lambda )\right] \left\vert {%
\partial _{{\nu }}\psi _{j}\left( \lambda \right) }\right\rangle .
\end{equation}%
Now we can derive the corresponding relation to Eq. (\ref{QGT2}) in the
non-Abelian case, \emph{i.e., }${Q_{\mu \upsilon }}={g_{\mu \upsilon }}-%
\frac{i}{2}{F_{\mu \upsilon }}$, where ${g_{\mu \upsilon }=}\left( {Q_{\mu
\upsilon }+Q_{\mu \upsilon }^{\dagger }}\right) /2$ and ${F_{\mu \upsilon }=i%
}\left( {Q_{\mu \upsilon }}-{Q_{\mu \upsilon }^{\dagger }}\right) $. Now all
the components have been extended to the matrix form, the symmetry and
anti-symmetry part correspond to the non-Abelian Riemannian metric and
non-Abelian Berry curvature, respectively.

\section{Geometric Characterizations of Ground-State
Properties}

We say that a QPT occurs if the \emph{qualitative} properties (no
matter local or topological) of the ground-state change when the
parameter of the Hamiltonian varies. However, it is hard to give an
explicit and general definition for in what degree a change of the
properties of the ground-state should be regarded as a
\emph{qualitative} change. The concept of the QGT
will shed light on this problem. 
Note that $dS$ represents the distance between two neighbor states located
on ${\lambda }$ and ${\lambda +d\lambda }$ of the Hamiltonian parameters
base manifold. Obviously, the quantity $\left( \Delta S/{\Delta \lambda }%
\right) ^{2}$ will be a good measurement about the degree of the
change in
ground-state respect to the change of the parameter. By using Eq. (\ref{ds2}%
) or (\ref{nads2}), we have
\begin{equation}
\lim_{{\Delta \lambda }\rightarrow 0}\frac{\Delta S^{2}}{{\Delta \lambda }%
^{2}}=\sum_{\mu ,\upsilon }\text{ }g_{\mu {\nu }}\text{ }\frac{{\partial
\lambda ^{\mu }}}{{\partial \lambda }}\otimes \frac{{\partial \lambda ^{\nu }%
}}{{\partial \lambda }}.  \label{FS}
\end{equation}%
By the way, we would like to point out that the scalar $\chi _{FS}:=\lim_{{%
\Delta \lambda }\rightarrow 0}\Delta S^{2}/{\Delta \lambda }^{2}$ in the $%
U(1)$ case provides a purely geometric definition for the
\emph{fidelity susceptibility }. \cite{Gu-PRE}
Note that $dS$ is the distance in the orthocomplement space of the subspace $%
\mathcal{H}_{E_{0}}(\lambda )$, so $\chi _{FS}$ is clearly not an intrinsic
geometric quantity for the induced fiber bundles.

However, there still exist an intrinsic geometric measurement. Consider the
parallel transport of ground-state around a small loop on the parameter base
manifold $\mathcal{M}$, there generally exist a $U(n)$ rotation of the
ground-state in the $nD$ internal subspace ${\psi }_{Final}=\mathcal{P}$exp$%
\left[ i\oint_{C}\mathcal{A}_{\mu }{d\lambda ^{\mu }}\right] {\psi }%
_{Initial}$, where $\mathcal{P}$ is the path-ordering operator and $\mathcal{%
A}_{\mu }$ is the Wilczek-Zee connection. In the first order approximation,
we have$\ \mathcal{P}$exp$\left[ i\oint_{C}\mathcal{A}_{\mu }{d\lambda ^{\mu
}}\right] \approx \boldsymbol{1}+i\sum_{\mu ,\upsilon }F_{\mu {\nu }}\sigma
^{\mu {\nu }}$, where $F_{\mu {\nu }}$ is the non-Abelian Berry curvature
and $\sigma ^{\mu {\nu }}$ is the area element ${d\lambda ^{\mu }\wedge
d\lambda ^{\nu }}$ of the infinitesimal surface $\sigma $, whose boundary is
the curve $C$. Now we can define a local geometric measurement depend on ${%
\lambda \in }\mathcal{M}$ as follows
\begin{equation}
\lim_{\sigma \rightarrow 0}\frac{\mathcal{P}exp\left[ i\oint_{C}\mathcal{A}%
_{\mu }{d\lambda ^{\mu }}\right] -\boldsymbol{1}}{\sigma }=\sum_{\mu
,\upsilon }F_{\mu {\nu }}\frac{{\partial \lambda ^{\mu }}}{{\partial \lambda
}}\wedge \frac{{\partial \lambda ^{\nu }}}{{\partial \lambda }}.
\end{equation}%
Note that the above definition is equivalent to $\left( {\psi }_{Final}-{%
\psi }_{Initial}\right) {\psi }_{Initial}^{-1}\sigma ^{-1}$ with
$\sigma \rightarrow 0,$ which is the measurement of the intensity of
relative change in the state along a infinitesimal loop. In compare
to Eq. (\ref{FS}), this is purely an instinct definition. We find
that the non-Abelian QGT make a
unification of the above two approaches through the relation $Q_{\mu {\nu }%
}=g_{\mu {\nu }}-iF_{\mu {\nu }}/2$. Note that the instinct curvature of the
induced fiber bundles ${F_{\mu \upsilon }=i}\left( {Q_{\mu \upsilon }}-{%
Q_{\mu \upsilon }^{\dagger }}\right) $ can be naturally associated
to the topological invariant Chern numbers, \emph{i.e., }the first
Chern number can be calculated as
\begin{equation}
C_{1}=\frac{i}{2\pi }\int_{\mathcal{M}}\left( {Q_{\mu \upsilon }}-{Q_{\mu
\upsilon }^{\dagger }}\right) {d\lambda ^{\mu }\wedge d\lambda ^{\nu }.}
\label{Chern number c1}
\end{equation}

Historically, a two-band model with a nonzero Chern number was first
proposed as by Haldane, \cite{Haldane} which was a honeycomb lattice
model with imaginary next-nearest-neighbor hopping. Haldane find the
two different symmetries breaking about the space reflection and the
time-reversal can be classified by the Chern numbers which reflect
the topology of the ground-state. \cite{Haldane2} Recently, Hatsugai
\cite{Hatsugai} proposed a non-Abelian Chern number for a ground
state multiplet with a possible degeneracy and found the system to
be a topological insulator when the energies of the multiplet are
well separated from the above.

Here we choose a two-band model as an example because it is one of the
simplest models that exhibit the topological
\begin{figure}[tbp]
\begin{center}
\includegraphics[width=3.3in]{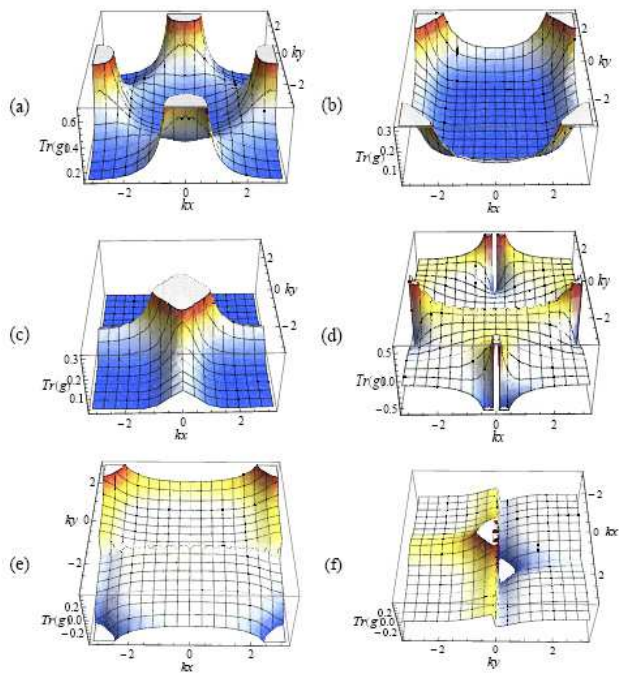} %
\end{center}
\caption{(color online) (a) The trace of the Riemannian metric $g$ as a
function of the $k_{x}$, $k_{y}$, in which the parameter $m=0$, (b) $m=2$,
(c) $m=-2$, (d) the Berry curvature $F$ as a function of the $k_{x}$, $k_{y}$%
, in which the parameter $m=0$, (e) $m=2$, (f) $m=-2$.}
\end{figure}
non-trivial states. This model was first introduced by Qi \cite{Qi}
and can be physically realized in
$\mathrm{Hg_{1-x}Mn_{x}Te/Cd_{1-x}Mn_{x}Te}$
quantum wells with a proper amount of $\mathrm{Mn}$ spin polarization. \cite%
{Liu}
The system under consideration is a two-dimensional (2D) lattice model on a two-torus $T^{2}$%
. The Hamiltonian is generally given by $H\left( k\right) =\varepsilon
\left( k\right) \boldsymbol{1}+\sum_{\alpha =1}^{3}d_{\alpha }\left(
k\right) \sigma ^{\alpha }$, where $\boldsymbol{1}$ is the $2\times 2$
identity matrix and $\sigma ^{\alpha }$ the three Pauli matrix. The
diagonalization of $H\left( k\right) $ is straightforward and the
eigenvalues can be written as $E_{\pm }(k)=\varepsilon \left( k\right) \pm
\sqrt{\sum_{\alpha =1}^{3}d_{\alpha }^{2}\left( k\right) }$ and eigenvectors
as $\Psi _{+}=\left( \cos \theta /2,e^{i\Phi }\sin \theta /2\right) ^{T}$ and $%
\Psi _{-}=\left( -\sin \theta /2,e^{i\Phi }\cos \theta /2\right) ^{T}$,
where $\theta =\arccos d_{3}\left( k\right) /\sqrt{d_{1}^{2}\left( k\right)
+d_{2}^{2}\left( k\right) +d_{3}^{2}\left( k\right) }$ and $\Phi =\arctan
d_{1}\left( k\right) /\sqrt{d_{1}^{2}\left( k\right) +d_{2}^{2}\left(
k\right) }$. In this model, the coefficients are given by $\varepsilon
\left( k\right) =0$, $d_{1}=\sin k_{x}$, $d_{2}=\sin k_{y}$ and $%
d_{3}=m+\cos k_{x}+\cos k_{y}$. In the thermodynamic limit, the
fiber can be chosen as the single-particle wave function and the
base manifold is the 2D momentum space
$\boldsymbol{k}=(k_{x},k_{y})$. Then QGT can be obtained by
substituting $\Psi _{-}$ into Eq. (\ref{QGT1}), we have
$Q_{xy}=\left(
\partial _{kx}\theta \partial _{ky}\theta +\partial _{kx}\Phi \partial
_{ky}\Phi \sin ^{2}\theta \right) /4+i\sin \theta \left( \partial _{kx}\Phi
\partial _{ky}\theta -\partial _{ky}\Phi \partial _{kx}\theta \right) /4$.
The corresponding Riemannian metric and Berry curvature can be obtained by
using the relation $g_{xy}=\Re Q_{xy}$ and $F_{xy}=-2\Im Q_{xy}$, we have $%
g_{xy}=\left( \partial _{kx}\theta \partial _{ky}\theta +\partial _{kx}\Phi
\partial _{ky}\Phi \sin ^{2}\theta \right) /4$ and $F_{xy}=-\sin \theta
\left( \partial _{kx}\Phi \partial _{ky}\theta -\partial _{ky}\Phi \partial
_{kx}\theta \right) /2$. We now analyze the behavior of the Riemannian
metric $g_{\mu {\nu }}$ and Berry curvature $F_{\mu {\nu }}$ when the
parameter $m$ is varied.
\begin{figure}[tbh]
\includegraphics[width=0.23\textwidth,height=0.13\textheight]{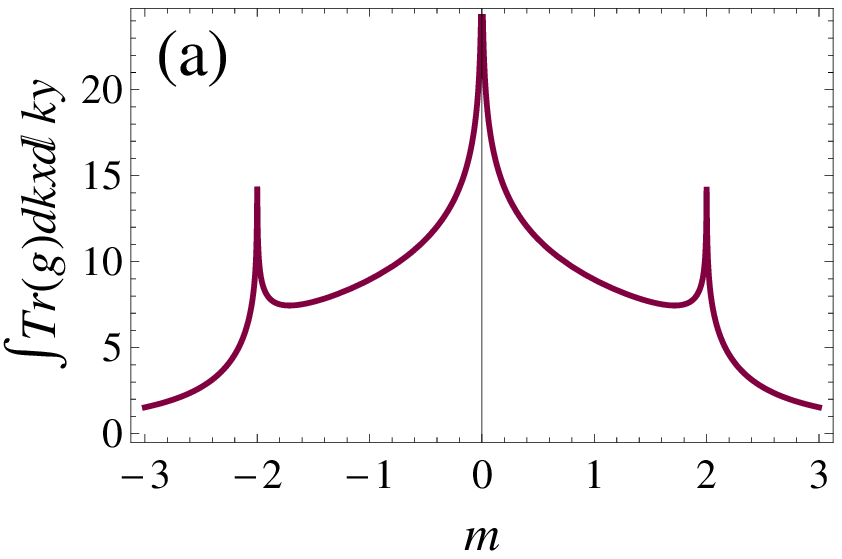} %
\includegraphics[width=0.23\textwidth,height=0.13\textheight]{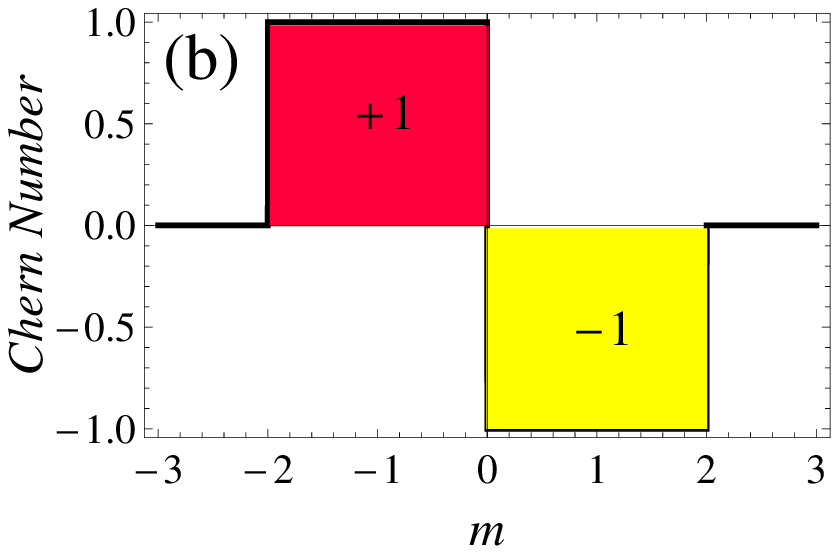}
\caption{(color online) (a) Integrate out $k_{x}$, $k_{y}$ of the Riemannian
metric $g$ as a function of $m$; (b) The first Chern number $c1$ as a
function of $m$.}
\end{figure}
The numerical results are presented in Fig. 1, in which the divergence
behavior are shown either in $g_{xy}$ or $F_{xy}$ at some region in $%
\boldsymbol{k}$-space when the parameter $m$ takes the values of -2, 0, 2.
This indicate a vanishing energy gap. The critical points can be clear shown
through integrating out $k_{x},k_{y}$ of the Riemannian metric $g_{xy}(m)$.
As shown in Fig. 2(a), we can see clearly a crown shape curve with three
divergent points at $m=-2$, $m=0,$ and $m=2$, respectively. In order to
distinguish the different topological phases, we can calculate the
topological invariant by using Eq. (\ref{Chern number c1})%
\begin{equation}
C_{1}=\left\{
\begin{array}{lll}
1 &  & \text{if}\quad -2<m<0 \\
-1 &  & \text{if}\quad 0<m<2 \\
0 &  & \quad \text{otherwise}%
\end{array}%
\right. ,
\end{equation}%
which is the first Chern number of the induced $U(1)$ line bundle. The phase
diagram is plotted in Fig. 2(b).

\section{Conclusions}

We have presented an adiabatic origin for the generalized QGT and
shown a non-Abelian QGT can naturally emerge from measuring the
distance between two neighbor states in the $U(N)$ vector bundle
induced by the adiabatic evolution of the quantum degenerate system.
In addition, we introduced two different local geometric
measurements based on an intuitive physical picture, i.e.,
$\lim_{{\Delta \lambda }\rightarrow
0}\Delta S^{2}/{\Delta \lambda }^{2}\ $ and $\lim_{\sigma \rightarrow 0}%
\left[ \mathcal{P}exp\left( i\oint_{C}\mathcal{A}_{\mu }{d\lambda ^{\mu }}%
\right) -\boldsymbol{1}\right] /\sigma $ to characterize the change
degree of the ground-state when the parameter varies. As the main
results of this work, we demonstrated the two different measurements
can be unified in the generalized non-Abelian QGT as its symmetric
and anti-symmetric parts, respectively. We showed the
symmetry-breaking QPTs and a class of topological QPTs can be
understood as the singular behavior of the local and topological
properties of the QGT in the thermodynamic limit. This approach shed
light on constructing a single framework to analyze QPTs with
topological orders.

\emph{Note added.}---\ After this manuscript has been submitted, we
noticed that a related work appeared, \cite{Rezakhani} which
similarly studies the geometry of quantum adiabatic evolution and
generalized the treatments of the quantum geometric tensor to allow
for degeneracy.

\section{Acknowledgments}

Y. Q. Ma thanks Shi-Jian Gu and Su-Peng Kou for helpful discussions.
This work was supported by the NSFC under Grants No. 10974234, No.
10821403, No. 10974247, No. 10874235, No. 10934010, and No.
60978019, 973 grant and National Program for Basic Research of MOST,
973 program under Grant No. 2010CB922904, the NKBRSFC under Grants
No. 2006CB921400, No. 2009CB930701, and No. 2010CB922904.

\end{document}